\documentclass{article}
\begin{document}
\begin{center}
{\large\bf Shape invariance and the
exactness of quantum Hamilton-Jacobi formalism}
\end{center}

\begin{center}{\large
Charles Cherqui\footnote{e-mail: ccherqu@luc.edu},  Yevgeny Binder\footnote{e-mail: yev@freeshell.org}, Asim Gangopadhyaya\footnote{e-mail: agangop@luc.edu}
}\\
Department of Physics, Loyola University Chicago, Chicago IL, USA
\end{center}
\begin{abstract}
Quantum Hamilton-Jacobi Theory and supersymmetric quantum mechanics (SUSYQM) are two parallel methods to determine the spectra of a quantum mechanical systems without solving the Schr\"odinger equation. It was recently shown that the shape invariance, which is an integrability condition in SUSYQM formalism, can be utilized to develop an  iterative algorithm to determine the quantum momentum functions. In this paper, we show that shape invariance also suffices to determine the eigenvalues in Quantum Hamilton-Jacobi Theory.
\end{abstract}

\vspace*{0.2in}
Quantum Hamilton-Jacobi Theory and supersymmetric quantum mechanics (SUSYQM)  are two very different methods that give eigenvalues for quantum mechanical systems without solving the Schr\"odinger differential eigenvalue equation.

Supersymmetric quantum mechanics is a generalization of Dirac's ladder operator method\footnote{Dirac attributes this method to Fock.} for the harmonic oscillator. This method consists of factorizing Schr\"odinger's second order differential operator into two first order differential operators that play roles analogous to ladder operators. If the interaction of a quantum mechanical system is described by shape invariant potentials  \cite{Cooper,Dutt}, SUSYQM allows one to generate all eigenvalues and eigenfunctions through algebraic methods.

Another formulation of quantum mechanics, the Quantum Hamilton-Jacobi (QHJ) formalism, was developed  by Leacock and Padgett  \cite{Leacock} and independently by Gozzi  \cite{Gozzi}.  It was made popular by a series of papers by Kapoor et. al.  \cite{Kapoor}.  In this formalism one works with the quantum momentum function (QMF) $p(x)$, which is related to the wave function $\psi$ through the relationship $p(x) = - \psi\prime(x)/\psi(x)$, where prime denotes differentiation with respect to $x$. Our definition of QMF's differs by a factor of $i\equiv \sqrt{-1}$ from that of ref.  \cite{Leacock,Kapoor,Constantin}, where they define $p_{\rm }(x)=-i\frac{\psi'(x)}{\psi(x)}$; we use $p_{\rm }(x)=-\frac{\psi'(x)}{\psi(x)}$.  It was shown, on a case by case basis, that the singularity structure of the function $p(x)$ determines the eigenvalues of the Hamiltonian  \cite{Leacock,Gozzi,Kapoor} for all known solvable potentials.  Kapoor and his collaborators have shown that the QHJ formalism can be used not only to determine the eigenvalues of the Hamiltonian of the system, but also its eigenfunctions  \cite{Kapoor2}. They have also used QHJ to analyze Quasi-exactly solvable systems where only an incomplete set of the eigenspectra can be derived analytically and also to study periodic potentials  \cite{Kapoor3}.  It is important to note that all cases worked out in Refs.   \cite{Leacock,Kapoor}  satisfied the integrability condition known as the translational shape invariance for which the SUSYQM method always gave the exact result. 

In an earlier paper  \cite{Constantin}, one of the authors (AG) and his collaborators had connected the Quantum Hamilton-Jacobi formalism to supersymmetry (SUSY) and shape invariance. Using shape invariance, it was shown that QMFs corresponding to different energies are connected via fractional linear transformations, and a general recursion formula was given for quantum momenta of any energy. Since QMFs are related to wave functions, Ref.  \cite{Constantin} provided an alternate recursive procedure to derive eigenfunction directly from shape invariance.

In this paper, we show that shape invariance also suffices to determine the singularity structure of the  QMFs, and hence provide an explanation why translational shape invariance also determines the eigenvalues of the system with the QHJ formalism.

In the QHJ formalism the spectrum of a quantum mechanical system is
determined by the solution of the equation:%
\begin{equation} %
p\,^2(x, \alpha_0
) -\,
p^{\,\prime}(x, \alpha_0
) = V(x, \alpha_0
) - E \label{QHJ1} \end{equation}%
Here we have chosen our units such that $\hbar=1$ and $2m=1$. 
$ \alpha_0
$ is a parameter characterizing the strength of the
potential. This equation is related to the Schr\"odinger equation
\begin{equation} -\psi^{\,\prime\prime} + \left( V(x, \alpha_0
)-E\right)\psi = 0
\label{SchEq1}
\end{equation}
via the correspondence
\begin{equation}%
p=- \left( \frac{\psi^{\,\prime}}{\psi}\right) ~{\rm whence
}~~~\psi(x) \sim e^{-\int p(x)dx}~~.\label{p-defined}
\end{equation}

\vspace*{.2in} \noindent In SUSYQM, the supersymmetric partner
potentials $V_\pm(x, \alpha_0
)$ are given by $W^2(x, \alpha_0
)\pm W^\prime(x, \alpha_0
)$ respectively. $W(x, \alpha_0
)$ is a real function, called the superpotential  \cite{Cooper}.  In this paper $V_-(x, \alpha_0
)$ is chosen as the potential for a Hamiltonian $H_-(x, \alpha_0
)$ whose eigenvalues we will be seeking. Thus, in terms of the superpotential $W(x, \alpha_0
)$, the Quantum Hamilton-Jacobi equation we will be solving is
\begin{equation} %
p\,^2(x, \alpha_0) -\,p^{\,\prime}(x, \alpha_0) = W^2(x, \alpha_0)- W^\prime(x, \alpha_0) - E \label{QHJ2} %
~.\end{equation}%
We will assume that our superpotential is such that SUSY remains unbroken for a range of values of the parameter $ \alpha_0
$. This implies that the ground state eigenvalue $E_0^{(-)}=0$  \cite{Cooper}, and the corresponding eigenfunction $\psi_0^{-}(x) \sim e^{-\int W(x, \alpha_0
)dx}$ is normalizable.  By differentiating $\psi_0^{(-)}(x)$, one can show that $W(x, \alpha_0) = - \, \left[  \psi_0^{(-)}(x) \right]^\prime /\psi_0^{(-)}(x)$. A comparison with Eq.  [\ref{p-defined}] shows that we should expect the following limit:%
\begin{equation}
 ~p\,(x) = W(x, \alpha_0)  ~~~~~~~~{\rm for}~~E=0.\label{Limit_of_p}%
\end{equation}
(Note that the partner Hamiltonian $H_+(x, \alpha_0)$, built using the potential $V_+(x, \alpha_0)$, has the same set of eigenvalues $E_n^{(+)} =E_{n+1}^{(-)}$, except for the groundstate.)

As stated earlier, in this paper our focus will be on $H_-(x, \alpha_0)$. In SUSYQM literature, it is customary to denote the eigenfunctions
of $H_-$ by $\psi^{(-)}$, and the corresponding QMF by $p_-(x) \equiv -\, \psi^{(-)\,\prime}/\psi^{(-)}$. Since we are going to be working with  $H_-$ and its solutions exclusively, we will henceforth suppress the ``-" subscript/superscript.

The QHJ method is centered around analyzing the pole structure of the quantum momentum function $p(x)$. Since the $n$-th eigenfunction of the Hamiltonian has $n$ nodes within the domain of the potential, the corresponding momentum function $p(x)$ will have $n$ singular points which can be shown to be simple poles  \cite{Kapoor}, each with a residue of $-1$. These are known as the ``moving poles".  Hence, a contour integral of $p(x)$ around the moving poles of the system yields:
\begin{eqnarray} %
\oint p(x) \,dx = - 2\pi i n ~.\label{Q_Condition0}
\end{eqnarray} %
This is the quantization condition for QHJ. For the sake of clarity, let us reiterate that the factor of $-i$ in the above quantization condition is due to our definition of the QMF differing from that of refs.  \cite{Leacock,Kapoor,Constantin} by a factor of $i$. Our choice is guided by our desire to keep $p(x)$ close in its definition to the superpotential $W(x)$ so Eq. (\ref{Limit_of_p})  holds, without any factor of $i$. For $n=0$, $p(x)$ would be identical to  $W(x)$.

The strength of QHJ derives from the fact that the contour of the above integration can be deformed to enclose ``fixed poles" instead (albeit traversing in the opposite direction  \cite{Leacock,Kapoor}). Thus, the quantization condition can now be written as%
\begin{eqnarray} %
\left. \oint p(x) \,dx \right|_{\rm origin} + \left. \oint p(x) \,dx \right|_{ \infty}= + 2\pi i n~~. \label{Q_Condition}
\end{eqnarray} %
The difference in sign between Eqs. (\ref{Q_Condition0}) and  (\ref{Q_Condition}) is due to the change in direction in which the integration is carried out. In this paper, we only consider systems that have fixed poles at the origin and/or $\infty$.

We now consider the impact of shape invariance on the Hamiltonian-Jacobi formalism.   Shape invariance implies the following relation among partner potentials of supersymmetric quantum mechanics:
\begin{eqnarray}  %
V_+(r, \alpha_0) = V_-(r, \alpha_1 ) + R(\alpha_0),\label{SIC_V}\nonumber%
\end{eqnarray}  %
i.e., potentials $V_+(r)$ and $V_-(r)$ have the same functional dependence on the coordinate variable $r$ and only differ from each other in the value of the parameter $ \alpha_0$ and an additive constant $R(\alpha_0).$ For all known shape invariant potentials, it is possible to express the constant $R(\alpha_0)$ as a difference:
\begin{equation}%
R(\alpha_0) = g( \alpha_1 ) - g(\alpha_0)~. \label{R_g}
\end{equation} %
We assume that $R(\alpha_0)$ can be written as a difference of $g$-functions for all shape invariant problems. 

The shape invariance condition, when written in terms of the superpotential $W$, takes the form:
\begin{equation}%
W^2(r, \alpha_0) + \frac{d\,W(r, \alpha_0)}{dr} = W^2(r, \alpha_1 ) -
\frac{d\,W(r, \alpha_1 )}{dr}+R(\alpha_0).\label{SIC_W}%
\end{equation} %
We will consider only the additive shape invariance with $ \alpha_1  = \alpha_0+1$, and use it to determine the structure of the superpotential near zero and $\infty$, two assumed end points of the domain.  It is essential that the knowledge of the superpotential at these boundary points be known in order to determine eigenvalues using QHJ. In particular, we will not consider cyclic shape invariant systems \cite{csis,csis2} because their superpotentials have an oscillatory behavior  at large distances; i.e, they are not well defined at infinity.

Here we shall divide all superpotentials into two categories: those with infinite domain $(0,\infty)$ or $(-\infty,\infty)$ and those with finite domain, such as $(x_1,x_2)$.  Superpotentials that are given in algebraic form or those in which a transformation such as $r = e^x$ converts it into an algebraic form will be considered in this paper. These potentials are defined over an infinite domain. We call them as ``Algebraic" and ``Hyperbolic" respectively. In the following, we will show how the shape invariance suffices to determine eigenvalues of generic algebraic and hyperbolic potentials. (We  will not be considering superpotentials with finite domain which we define as trigonometric.)

\vspace*{.25in}
\noindent
{\large\bf Algebraic Potentials:} \\In this category, we assume that the potential is explicitly expressed in an algebraic form.  Let us assume the  superpotential has the following structure near origin:%
\begin{eqnarray}  %
W(r, \alpha_0) &=& \sum_{k=-1}^{\infty} b_{k}(\alpha_0)\,r^k~=~ \frac{b_{-1}(\alpha_0)}{ r} \,+\, b_{0}(\alpha_0)\, r^{0}
\,+\, b_{1}(\alpha_0) \,r^{1} \,+\, b_{2}(\alpha_0) \,r^{2}
\,+\,\cdots   \nonumber %
\end{eqnarray} 
We have not included a term more singular than $1/r$ in the superpotential to avoid the particle's ``falling to the center." Particles cannot be sustained in a bound state in an attractive singular potential if the singularity is higher than $-\frac1{4r^2}~~$ \cite{SingularPotential}.  Additionally, we are interested in the simple poles  of the QMF at boundary points, which are expected to be related to the same order singularities of the  superpotential at those points.

The derivative of the superpotential is then given by
\begin{eqnarray}  %
\frac{d\,W(r, \alpha_0)}{dr} &=& -\frac{b_{-1}(\alpha_0)}{ r^2} \,+\,
0 \,+\,
b_{1}(\alpha_0)  \,+\, 2\, b_{2}(\alpha_0) \,r \,+\,\cdots   \nonumber ~~;%
\end{eqnarray}  and the square of the superpotential is given by
\begin{eqnarray}  %
W^2(r, \alpha_0)
&=& \left( b_{-1} (\alpha_0) \right)^2 r^{-2} + %
\left( 2\, b_{-1} (\alpha_0) ~b_{0} (\alpha_0) \right) r^{-1} + %
\left( \left( b_{0} (\alpha_0) \right)^2 + 2\, b_{-1} (\alpha_0) ~b_{1} (\alpha_0) \right)  + \cdots%
\nonumber\end{eqnarray}  %
Substituting these expansions into the shape invariance condition (\ref{SIC_W}), we arrive at the following constraints on coefficients of the superpotential $W(x, \alpha_0)$:
\begin{eqnarray} %
\begin{array}{ll}
{\rm from ~the}~r^{-2}~{\rm terms}  ~~~~~~~~  ~~~~~~~~  ~~~~~~~~ &\left( b_{-1} (\alpha_0) \right)^2 - {b_{-1}(\alpha_0)}  %
=%
\left( b_{-1} ( \alpha_1 ) \right)^2 + {b_{-1}( \alpha_1 )}
\\
{\rm from ~the}~r^{-1}~{\rm terms}  ~~~~~~~~ &\left( 2\, b_{-1} (\alpha_0) ~b_{0} (\alpha_0) \right) %
=%
\left( 2\, b_{-1} ( \alpha_1 ) ~b_{0} ( \alpha_1 ) \right)
\\
{\rm from ~the}~r^{0}~{\rm terms}  ~~~~~~~~ &\left( \left( b_{0} (\alpha_0) \right)^2 + 2\, b_{-1} (\alpha_0)
~b_{1} (\alpha_0) \right)  + b_{1}(\alpha_0) \\
&=%
\left( \left( b_{0} ( \alpha_1 ) \right)^2 + 2\, b_{-1} ( \alpha_1 )
~b_{1} ( \alpha_1 ) \right)  -  b_{1}( \alpha_1 ) +  R(\alpha_0)  \label{constraint_b}%
\end{array}\end{eqnarray}  %
The first of these three constraints of Eq. (\ref{constraint_b}) is a quadratic equation for the coefficient $b_{-1} ( \alpha_1 )$ in terms of $b_{-1} (\alpha_0)$. It has two solutions: ${b_{-1}( \alpha_1 )} = -\, {b_{-1}(\alpha_0)}$ ~or~ ${b_{-1}( \alpha_1 )} = {b_{-1}(\alpha_0)} - 1$.  Since $b_{-1}(\alpha_0)$ is the coefficient of the most dominant term and hence determines the structure near the origin, its value plays an essential role in deciding whether SUSY is unbroken%
\footnote{The limiting values of the superpotential at two boundary points of the domain determine whether SUSY is broken or unbroken. If the value of the superpotential at both end points have the same sign SUSY is broken and SUSY is unbroken if both end points have opposite signs.  Thus, if the sign of $W$ at one end is altered relative to the other end, the SUSY goes from being unbroken to broken or  vice versa. For more details, see Ref. \cite{Cooper,Dutt}}.  %
Hence, the first solution ${b_{-1}( \alpha_1 )} = -\, {b_{-1}(\alpha_0)}$, is not acceptable if both $b_{-1}( \alpha_0)$ and ${b_{-1}(\alpha_1)}$ were to keep the system in a parameter domain needed for unbroken SUSY. Thus, the first constraint  (\ref{constraint_b}) implies the relationship %
${b_{-1}( \alpha_1 )} = {b_{-1}(\alpha_0)} - 1$. For $ \alpha_1 = \alpha_0+1$, this gives  the
difference equation ${b_{-1}( \alpha_0+1)} = {b_{-1}(\alpha_0)}- 1$,
whose solution is given by  %
\begin{eqnarray} 
b_{-1}(\alpha_0) = - \alpha_0 + {\rm constant}; \label{b_1}%
\end{eqnarray} 
we choose the ${\rm constant}=0$. This choice, while it simplifies the calculation, has no effect on the final result. The second constraint leads to the difference equation%
\begin{eqnarray} 
\alpha_0 \, b_{0} (\alpha_0) =  \alpha_1  \,b_{0} ( \alpha_1 )~. \label{b_0}%
\end{eqnarray} 
In other words, the product $\alpha_0\,b_{0} (\alpha_0)$ does not depend
on the parameter $\alpha_0$. Hence, we have $$b_{0} (\alpha_0) =
\frac{\beta}{\alpha_0} ~~,$$ where $\beta$ is a constant independent of the parameter $ \alpha_0$. Thus, near the origin, the structure of the superpotential $W(x, \alpha_0)$ is given by:
\begin{eqnarray}  %
\left.  W\left(r\,,\, \alpha_0\right)\right|_{\rm origin}= -\,\frac{\alpha_0}{ r} \,+\,
\frac{\beta}{\alpha_0} \,+\,\cdots  \label{W_origin}%
\end{eqnarray} 
Now, let us explore the structure of the superpotential near $\infty$.  To do this, we
define a variable $u=\frac1r$, and $W\left(1/u\,,\, \alpha_0\right) \equiv  \widetilde{W}\left(u\,,\, \alpha_0\right)$. The shape invariance condition of Eq.  (\ref{SIC_W}) thus transforms to
\begin{equation}%
\widetilde{W}^2\left(u\,,\, \alpha_0\right) -u^2 ~
\frac{d\,\widetilde{W}\left(u\,,\, \alpha_0\right)}{du} =
\widetilde{W}^2\left(u\,,\, \alpha_1 \right) + u^2  ~
\frac{d\,\widetilde{W}\left(u\,,\, \alpha_1 \right)}{du}+R(\alpha_0).%
\label{SIC-transformed}
\end{equation} %
Analogous to the expansion near the origin, let us expand $W\left(\frac1u\,,\, \alpha_0\right)$ in powers of $u$
as we have done earlier.
\begin{eqnarray}  %
\widetilde{W}\left(u\,,\, \alpha_0\right) &=& \sum_{k=-1}^{\infty} c_{k}(\alpha_0)\,u^k ~=~  \frac{c_{-1}(\alpha_0)}{ u} \,+\, c_{0}(\alpha_0)\, u^{0}
\,+\, c_{1}(\alpha_0) \,u^{1} \,+\, c_{2}(\alpha_0) \,u^{2}
\,+\,\cdots   \nonumber %
\end{eqnarray} 
\begin{eqnarray}  %
\widetilde{W}^2\left(u, \alpha_0\right) &=&
\frac{ \left(c_{-1} (\alpha_0) \right)^2}{u^{2}} + %
 \frac{ \left( 2\, c_{-1} (\alpha_0) ~c_{0} (\alpha_0) \right)}{u} + %
\left( \left( c_{0} (\alpha_0) \right)^2 + 2\, c_{-1} (\alpha_0) ~c_{1} (\alpha_0) \right)  + \cdots \nonumber %
\end{eqnarray} 
\begin{eqnarray}  %
\frac{d\,{W}(r, \alpha_0)}{dr} =
-u^2~\frac{d\,\widetilde{W}(u, \alpha_0)}{du}%
&=& {c_{-1}(\alpha_0)}  \,-\,
c_{1}(\alpha_0)\,u^2  \,-\, 2\, c_{2}(\alpha_0) \,u^3 \,+\,\cdots    \nonumber~~;%
\end{eqnarray} 
Substituting these expressions in the transformed shape invariance condition of Eq. (\ref{SIC-transformed}), we get %
\begin{eqnarray}  %
&{\rm from ~the}~r^{-2}~{\rm terms}  ~~~~~~~~&\left(c_{-1} (\alpha_0) \right)^2  = \left( c_{-1} ( \alpha_1 )
\right)^2 \label{constraint_c1}~~;\\
&{\rm from ~the}~r^{-1}~{\rm terms}  ~~~~~~~~&\left( 2\, c_{-1} (\alpha_0) ~c_{0} (\alpha_0) \right)  %
\left( 2\, c_{-1} ( \alpha_1 ) ~c_{0} ( \alpha_1 ) \right) \label{constraint_c2}~;~~~~~~~{\rm and}\\
&{\rm from ~the}~r^{0}~{\rm terms}  ~~~~~~~~&\left( c_{0} (\alpha_0) \right)^2 + 2\, c_{-1} (\alpha_0) ~c_{1} (\alpha_0) +{c_{-1}(\alpha_0)} \nonumber\\&& %
=%
\left( c_{0} ( \alpha_1 ) \right)^2 + 2\, c_{-1} ( \alpha_1 ) ~c_{1}
( \alpha_1 )-{c_{-1}( \alpha_1 )}  + R(\alpha_0)  \label{constraint_c3}
~. \end{eqnarray} 
From Eq.  (\ref{constraint_c1}), we get $c_{-1} ( \alpha_1 ) = \pm \,c_{-1} (\alpha_0)$. As stated earlier we assumed that SUSY remains unbroken for a range of values of the parameter, and since $c_{-1} ( \alpha_1 ) = - \,c_{-1} (\alpha_0)$ would lead to breaking of SUSY, we choose $c_{-1} ( \alpha_1 ) = \,c_{-1} (\alpha_0)=c_{-1}$; i.e., this coefficient does not depend on the parameter $ \alpha_0$.  This then further leads to $c_{0} (\alpha_0)=c_{0} ( \alpha_1 )=c_{0}$, another constant.  Eq. (\ref{constraint_c3}) now yields, %
$2\, c_{-1}  ~c_{1} (\alpha_0) + c_{-1} = %
2\, c_{-1} ~c_{1} ( \alpha_1 )-c_{-1} + g(\alpha_1) - g(\alpha_0) .$ Which gives %
\begin{eqnarray} 
c_{1} ( \alpha_1)\,+ \frac{g(\alpha_1)}{2\,c_{-1}}= \, c_{1} (\alpha_0) +\frac{g(\alpha_0)}{2\,c_{-1}} + 1 ~. \label{eq19}
\end{eqnarray} %
We can write this equation as
\begin{eqnarray} 
f( \alpha_1)= f(\alpha_0) + 1 ~, \label{eq20}
\end{eqnarray} %
where $f(\alpha_0) = c_{1} ( \alpha_0)\,+ \frac{g(\alpha_0)}{2\,c_{-1}}$ . Since $\alpha_1 = \alpha_0+1$, the solution of Eq. (\ref{eq20}) is $f(\alpha_0) = \alpha_0 +\Delta$. Hence, we have
 $c_{1} (\alpha_0) = \alpha_0 -\frac{g(\alpha_0)}{2\,c_{-1}} + \Delta$, where $\Delta$ is a constant independent of $ \alpha_0$. Thus, near $\infty$ the superpotential is given by
\begin{eqnarray}  %
\left. \widetilde{W}\left( u\,,\, \alpha_0\right) \right|_{\infty}= \frac{c_{-1}}{ u} \,+\, c_{0}
\,+\, \left( \alpha_0-\frac{g(\alpha_0)}{2\,c_{-1}}  +\Delta\right) \,u \,+\, \cdots  \label{W_at_Infinity4}%
\end{eqnarray}  %
The result we have obtained depended crucially upon the assumption that ${c_{-1}}$ is not zero.  If ${c_{-1}}=0$, the structure of the potential near $\infty$ will be very different and the constraints given by Eqs.  \mbox{(\ref{constraint_c1})-(\ref{constraint_c3})} would no longer be valid. We would need to do the calculations again. This case is presented next.

For ${c_{-1}}=0$, the superpotential near $\infty$ is given by  $\widetilde{W}(u, \alpha_0) = c_0 + c_1u + c_2u^2 + \cdots$. This leads to $\widetilde{W}^2(u, \alpha_0) = c_0^2 + 2c_0c_1u + (c_1^2 + 2c_0c_2)u^2 + \cdots$, and $u^2 \frac{d\widetilde{W}}{du} = c_1u^2 + 2c_2u^3 + \cdots$.
Since $\widetilde{W}^2(u, \alpha_0)$ and $ \widetilde{W}^2(u,  \alpha_1 )$  must satisfy the shape invariance condition (\ref{SIC-transformed}), matching the first two powers of $u$ we get
\begin{equation}\begin{array}{ll}
{\rm from ~the}~r^{0}~{\rm terms}  ~~~~~~~~ ~~~~~~~~& c_0^2( \alpha_1 ) = c_0^2( \alpha_1 ) + R(\alpha_0) \label{coul_case2_u0} \\
{\rm from ~the}~r^{1}~{\rm terms}  ~~~~~~~~ ~~~~~~~~&  c_0(\alpha_0)c_1(\alpha_0) = c_0( \alpha_1 )c_1( \alpha_1 ) \label{coul_case2_u1}.
\end{array}
 \end{equation}
The first constraint of Eq. (\ref{coul_case2_u0}) may be rewritten using (\ref{R_g}) as $c_0^2(\alpha_0) + g(\alpha_0) = c_0^2(\alpha_0 + 1) + g(\alpha_0+ 1)$ which means that the quantity $c_0^2(\alpha_0) + g(\alpha_0)$ is independent of the argument $\alpha_0$, and hence equal to a constant $\Lambda$. This leads to
\begin{eqnarray}  c_0(\alpha_0) = \pm \sqrt{-g(\alpha_0) + \Lambda}~. \label{c_0} \end{eqnarray} 
Now from the second constraint we see  that  the product $c_0(\alpha_0)c_1(\alpha_0)$ is also independent of the argument $\alpha_0$, and hence it must also be equal to a constant, which we denote by $\zeta$.  Thus, from $c_0(\alpha_0)c_1(\alpha_0)=\zeta$, we have
\begin{eqnarray}  c_1(\alpha_0) = \frac{\zeta}{c_0 (\alpha_0)} = \frac{\zeta}{\pm \sqrt{-g(\alpha_0) + \Lambda}}. \label{c_1} \end{eqnarray} 
Thus, near $\infty$,
\begin{eqnarray}  \widetilde{W}(u, \alpha_0) =\left[\pm \left(  \sqrt{-g(\alpha_0) + \Lambda} \right) ~\pm \left(  \frac{\zeta}{\sqrt{-g(\alpha_0) + \Lambda}}\right)~u +\cdots\right]\label{W_infty_2} \end{eqnarray} 
Thus, we have gathered significant knowledge about the structure of the superpotential near boundary points simply from the shape invariance. Recapitulating:
near the origin, as given in Eq.  (\ref{W_origin}),  the structure is
$  W\left(r\,,\, \alpha_0\right)= -\,\frac{\alpha_0}{ r} \,+\,\frac{\beta}{\alpha_0} \,+\,\cdots $.
Near $\infty$, as $u \rightarrow 0$, there are two possible structures that depend on whether the superpotential has a singularity; i.e., whether the value of the coefficient $ \,c_{-1} (\alpha_0)$ is zero or non-zero. They are:
\begin{eqnarray} 
\widetilde{W}(u, \alpha_0) =
    \left\{
        \begin{array}{ll}
		\frac{c_{-1}}{ u} \,+\, c_{0} \,+\, \left( \alpha_0-\frac{g(\alpha_0)}{2\,c_{-1}}  +\Delta\right) \,u \,+\, \cdots
		                & ~~~~~c_{-1} (\alpha_0)   \neq 0 \\
		                &\\
		\pm  \sqrt{-g(\alpha_0) + \Lambda} ~\pm \frac{\zeta}{\sqrt{-g(\alpha_0) + \Lambda}}~u +\cdots
		                & ~~~~~c_{-1} (\alpha_0)   = 0.
	\end{array}
    \right.
\label{c-negative-1}
\end{eqnarray} 

With the understanding about the structure of $W(x, \alpha_0)$ at the end points of the domain, we now substitute these expansions of $W$ into  QHJ and obtain the implications on the QMFs, and hence on the eigenvalues of the hamiltonian.  We first expand the momentum $p(r)$ near the origin as:
\begin{eqnarray}  %
p(r, \alpha_0)  &=& \sum_{k=-1}^{\infty} p_{k}\,r^k = \frac{p_{-1} }{ r} \,+\, p_{0} \, r^{0} \,+\, p_{1}
\,r^{1} \,+\, p_{2}  \,r^{2}
\,+\,\cdots   \nonumber %
\end{eqnarray} 
The derivative of the momentum $p$ is then given by
$\frac{d\,p(r, \alpha_0)}{dr} = -\frac{p_{-1} }{ r^2} \,+\, 0 \,+\,
p_{1}   \,+\, 2\, p_{2}  \,r \,+\,\cdots   $
and the square of $p$ is given by
%
$p^2(r, \alpha_0)
= \frac{\left( p_{-1}   \right)^2}{r^{2}} + %
\frac{\left( 2\, p_{-1}\,p_{0}   \right)}{r} + %
\left( p_{0}^2 + 2\, p_{-1}\,p_{1}   \right)  + \cdots$.%

Collecting terms with various powers of $r$, we find that near the origin the combination $p^2 - p'$ is given by
\begin{eqnarray} 
\left.  \left(  p^2 - p' \right)  \right|_{\rm origin}  \sim \frac{ p_{-1}^2 + p_{-1}}{r^{2}} + %
\frac{\left( 2\, p_{-1}\,p_{0}   \right)}{r} + %
\left( p_{0}^2 + 2\, p_{-1}\,p_{1}  - ~p_{1} \right) +
\cdots \nonumber%
 \end{eqnarray} 
Substituting into the QHJ equation (\ref{QHJ2}) near $r = 0$, we
get %
\begin{eqnarray} %
p_{-1} = -\alpha_0~~.
\end{eqnarray} 
It is this term in the expansion of $p$ that contributes to the contour integration around the origin in the complex $r$ plane. So, we do not need other coefficients.

To determine the leading order behavior of the momentum function at $\infty$, let us first consider the case that $c_1(\alpha_0) \neq 0$. We expand the momentum function $p\,\left(\frac1u\,,\, \alpha_0\right)\equiv {\widetilde p}\left(u\,,\, \alpha_0\right)$ in powers of $u$:
\begin{eqnarray}  %
{\widetilde p}\left(u\,,\, \alpha_0\right) &=& \sum_{k=-1}^{\infty} {\widetilde p}_{k} \,u^k = \frac{{\widetilde p}_{-1} }{ u} \,+\, {\widetilde p}_{\,0} \,
u^{0} \,+\, {\widetilde p}_{1}  \,u^{1} \,+\, {\widetilde p}_{2} \, u^{2}
\,+\,\cdots   \nonumber %
\end{eqnarray} 
This leads to the following expression for  the combination $\widetilde{p}^2 - \widetilde{p}\,'$:
\begin{eqnarray} 
\left.  \left(  \widetilde{p}^2 - \widetilde{p}' \right)  \right|_{\infty} =%
\left.\left( \widetilde{p}^2 + u^2 \frac{d\widetilde{p}}{du}\right)\right|_{\infty} %
\sim {\widetilde p}_{-1}^{\,2} u^{-2} + %
\left( 2\, {\widetilde p}_{-1} \,{\widetilde p}_{0}   \right) u^{-1} + %
\left( {\widetilde p}_{0}^{\,2} + 2\, {\widetilde p}_{-1}\,{\widetilde p}_{1} -%
{{\widetilde p}_{-1} }   \right)  + \cdots%
\end{eqnarray} 

Substituting into Eq.  (\ref{QHJ2}), 
$\widetilde{p}^2 + u^2 \frac{d\widetilde{p}}{du}  = \widetilde{W}^2+u^2\frac{d\widetilde{W}}{du} -E~,$ 
we obtain %
\begin{eqnarray} %
\frac{\left( \widetilde{p}_{-1}\right)^2 }{u^2 } +
\frac{2\widetilde{p}_0\widetilde{p}_{-1}}{u} + %
\left(\widetilde{p}_0^2 + 2\, \widetilde{p}_{-1} \,\widetilde{p}_1  -
\widetilde{p}_{-1} \right) + \cdots &=& \frac{\left( c_{-1}\right)^2 }{u^2 } + \frac{2c_0c_{-1}}{u}  \\
&&+ \left( c_0^2 + 2\, c_{-1} \left( \alpha_0 - \frac{g(\alpha_0)}{2\,
c_{-1}} \right) - c_{-1} -E\right)~. \nonumber%
\end{eqnarray} %

\vspace*{0.2in}
\noindent
By matching the various powers of the variable $u$, we get
$\widetilde{p}_{-1}= c_{-1} ~;~\widetilde{p}_{0}= c_{0}$;  and
$ \widetilde{p}_0^2 + 2\, \widetilde{p}_{-1} \,\widetilde{p}_1  - %
\widetilde{p}_{-1}=c_0^2 + 2\, c_{-1} \left( \alpha_0 - \frac{g(\alpha_0)}{2\,
c_{-1}} \right) - c_{-1} -E$.
From the last equality, we get $ 2\, c_{-1} \left( \alpha_0 -
\frac{g(\alpha_0)}{2\, c_{-1}}\right)   -E= \left( 2\, c_{-1}
\widetilde{p}_1   \right)$, i.e., %
$\widetilde{p}_1 = \alpha_0 - \frac{g(\alpha_0)+E}{2\, c_{-1}}.$ At $\infty$,  it is this last coefficient  $\widetilde{p}_1$ that contributes to the contour integration.

We are now ready to determine the eigenvalues for the system using
the information we have generated solely from shape invariance.
The contribution to the contour integration is given by
\begin{eqnarray} 
\left[  \oint p \,dr \right]_{\rm Moving~poles} &=& - \oint_{r=0} p \, dr - \oint_{r=\infty} p \, dr \nonumber \\&&\nonumber\\
&=&  - \oint_{r=0} p \, dr  + \oint_{u=0} \frac1{u^2} ~\widetilde{p} \, du \nonumber \\
&&\nonumber \\
&=& -(2\pi i p_{-1}) +  (2\pi i  \widetilde{p}_1) \nonumber\\
&&\nonumber \\
 &=& - 2\pi \,i \left(-\alpha_0\right)  +   2\pi \,i \left[\alpha_0 -
\frac{g(\alpha_0)+E}{2\, c_{-1}} \right]~  \nonumber\\
&&\nonumber \\
 &=&  2\pi \,i \left[2\alpha_0 -
\frac{g(\alpha_0)+E}{2\, c_{-1}} \right]
~~.\label{Q_Condition1}
\end{eqnarray} 
Please note that due to the $\frac1{u^2}$ factor in the term $\oint_{u=0} (\frac1{u^2} ~\widetilde{p}) \, du$, only the linear term of $\widetilde{p}$ contributes towards the integral and is given by $(2\pi i  \widetilde{p}_1)$.

It is worth noting that we cannot get the entire global behavior of the potential from expansions at the origin and $\infty$. If the potential is a symmetric function of $r$, as is the case for the three dimensional harmonic oscillator, one gets an additional contribution \cite{Leacock,Kapoor}. As one embeds the domain $(0,r)$ in a two-dimensional complex plane, a mirror image of the potential function in the region ${\cal R}eal(r)<0$ also generates singularities for $p(r)$ exactly at the same points as $p(r)$ does in the region ${\cal R}eal(r)>0$.  As a consequence, we get
\begin{eqnarray} 
\left[  \oint p \,dr \right]_{\rm Moving~poles} \equiv \left[  \oint p \,dr \right]_{{\cal R}eal(r)>0}&=& - \left[  \oint p \,dr \right]_{{\cal R}eal(r)<0}  - \oint_{r=0} p \, dr  + \oint_{u=0} \frac1{u^2} ~\widetilde{p} \, du
\nonumber\end{eqnarray} 

\vspace*{.1in}
\noindent
However, $\left[  \oint p \,dr \right]_{{\cal R}eal(r)>0}$ and $\left[  \oint p \,dr \right]_{{\cal R}eal(r)<0} $ have identical value due to the symmetry of the potential. This leads to
\vspace*{.1in}
\begin{eqnarray} 
\left[  \oint p \,dr \right]_{\rm Moving~poles} &=& \frac12\left( - \oint_{r=0} p \, dr  + \oint_{u=0} \frac1{u^2} ~\widetilde{p} \, du\right) = -2\pi \,i\,n~.
\end{eqnarray} 
From Eq. (\ref{Q_Condition1}), we then get %
$$\frac12\, \left\{  2\pi \,i \left[2\alpha_0 - \frac{g(\alpha_0)+E_n}{2\, c_{-1}} \right] \right\}= -2\pi i n~.  $$
Solving for the energy, the Eq.  (\ref{Q_Condition1})  gives $E_n = 4\, c_{-1}\left(n+ \alpha_0  \right) -g(\alpha_0) .$ To determine the value of $g(\alpha_0) $, we remember that for unbroken SUSY that the groundstate energy $E_0 =0$. Thus, substituting for $n=0$, we get $g(\alpha_0) = 4\, c_{-1}\alpha_0$, and hence $g(\alpha_n) = 4\, c_{-1}\left(\alpha_0  +n\right)$.
Substituting for $g(\alpha_0)$ and $g(\alpha_n) $ in $E_n$, we get
\begin{eqnarray} %
E_n = 4\,n\, c_{-1}~ = ~ g(\alpha_n) - g(\alpha_0).\label{3D-result1}%
\end{eqnarray} %
This is the desired result for any shape invariant problem.
For the $3$-dimensional harmonic oscillator, one would identify the coefficient $ c_{-1}$ with $\frac12\, \omega$, and the energy would be given by
\begin{eqnarray} %
E_n = 4\,n\, c_{-1}~= 2 n \omega~,\label{3D-result2}%
\end{eqnarray} %
as expected. However, the result of Eq. (\ref{3D-result1}) is valid for any superpotential that is a) shape invariant  and b) has singularities at both infinity and the origin.

Now let us consider the case that the superpotential has no divergence at $\infty$, i.e., $c_{-1} = 0$. For this case, we will need to determine the structure of the quantum momentum function $p(x)$ at $\infty$ (the structure at the origin remains the same as that for ${c_{-1} \neq 0}$.) We substitute the expansion of superpotential $\widetilde{W}(u, \alpha_0) $ from Eq.  (\ref{c-negative-1}) into the Quantum Hamilton-Jacobi equation at $\infty$, and we get
\begin{eqnarray} %
\frac{\left( \widetilde{p}_{-1}\right)^2 }{u^2 } +
\frac{2\widetilde{p}_0\widetilde{p}_{-1}}{u} + %
\left(\widetilde{p}_0^2 + 2\, \widetilde{p}_{-1} \,\widetilde{p}_1  -
\widetilde{p}_{-1} \right) + \cdots %
&=& %
{(-g(\alpha_0) + \Lambda - E)} ~+ 2\zeta~u +\cdots
~. \nonumber%
\end{eqnarray} %
Now equating the coefficients of various powers of $u$, we get ~$\widetilde{p}_{-1}(\alpha_0)=0$;
$\widetilde{p}_0^2(\alpha_0) =  -g(\alpha_0) + \Lambda -E$ ;
$\widetilde{p}_0(\alpha_0)\widetilde{p}_1(\alpha_0) = \zeta$;
etc.
From these, we obtain the following coefficients for $p$:
$$\widetilde{p}_0 =  \pm \sqrt{-g(\alpha_0) + \Lambda - E} ;~~~~{\rm and}~~~~\widetilde{p}_1 = \pm \frac{\zeta}{\sqrt{-g(\alpha_0) + \Lambda - E}}~.$$
The quantization condition yields:
\begin{eqnarray} 
\left[  \oint p \,dr \right]_{\rm Moving~poles} &=&
 - \oint_{r=0} p \, dr  + \oint_{u=0} \frac1{u^2} ~\widetilde{p} \, du \nonumber \\
&&\nonumber \\second
&=& -(2\pi i p_{-1}) +  (2\pi i  \widetilde{p}_1) \nonumber\\
 &=& - 2\pi \,i \left(-\alpha_0\right)   +   2\pi \,i \left[\pm \frac{\zeta}{\sqrt{-g(\alpha_0) + \Lambda - E}} \right]~  \nonumber\\
 &=&  -2\pi\,i\,n \,~~.\label{Q_Condition2}
\end{eqnarray} 
Thus, we have $-\alpha_0 - \frac{\zeta}{\pm \sqrt{-g(\alpha_0) + \Lambda - E}} =  n$.
Solving for $E_n$,
\begin{eqnarray}  E_n = \frac{\zeta}{(n+\alpha_0)^2} + g(\alpha_0) - \Lambda. \label{energy2} \end{eqnarray} 
Since we are in the domain of unbroken SUSY, we must have $E_0 = 0$. Hence substituting $n=0$ in Eq.  (\ref{energy2}), we get	 $g(\alpha_0) = - \frac{1}{\alpha_0^2} + \Lambda$; i.e.,  $ g(\alpha_n) = - \frac{1}{(\alpha_0 + n)^2} + \Lambda$.  The energy $E_n$ is then given by
\begin{eqnarray} 
E_n = \zeta^2 \left[ \frac{1}{(\alpha_0)^2} - \frac{1}{(n+\alpha_0)^2}\right] = g(\alpha_n) -g(\alpha_0). \label{result-expo-W}
\end{eqnarray} 
Thus, we determined the energy of this system described by a potential that is singular at the origin and has no singularity at $\infty$ (Coulomb-like) simply by employing the shape invariance condition. Once we identify $\zeta^2$ with $e^2/2$ and $\alpha_0$ with $l+1$, this energy is exactly the same as that for the Coulomb potential. Once again, we state that Eq. (\ref{result-expo-W}) is valid not only for the Coulomb potential, but for any shape invariant potential that has a fixed pole at the origin and none at the $\infty$.

\vspace*{.5in}
\noindent
{\bf \large Hyperbolic potentials:\\}
Now we consider superpotentials that arise as a linear combination of various powers of a basic exponential function $e^{x}$. For example, the Morse, Eckart, hyperbolic Rosen-Morse potentials would fall under this category. We assume that our generic superpotential has the general form
\begin{eqnarray} 
W(x, \alpha_0) = \sum_j b_j \left[  e^{x}\right]^j. \label{expo-W}
\end{eqnarray} 
This superpotential has singularities at $\pm \infty$.  To analyze the nature of these singularities we first convert it into an algebraic form with a transformation $r=e^{x}$. This  change of the variable maps $x\in (-\infty,\infty)$ to $r\in (0,\infty)$. We then subject the resulting superpotential to the shape invariance condition.

Near negative $\infty$;  i.e., $x \rightarrow -\infty$, and $r \rightarrow 0$, we assume the potential has the following structure:
\begin{eqnarray} 
W(x, \alpha_0) = b_{-1}r^{-1} + b_0 + b_{1}r + \cdots \label{exp-Wr}.
\end{eqnarray} 
Near positive $\infty$;  $r \rightarrow \infty$, we express it in terms of a variable $u=1/r = e^{-x}$, where we assume it has the following structure:
\begin{eqnarray} 
\widetilde{W}(u, \alpha_0) = c_{-1}u^{-1} + c_0 + c_{1}u + \cdots \label{exp_Wu}.
\end{eqnarray} 
Now we will determine how shape invariance constrains various terms in these two expansions at the end points of the domain of $r$.

As with the algebraic superpotential, the following procedure differs depending on whether $b_{-1}$ and $c_{-1}$ are each zero or nonzero, making for four possible cases. We want to stress that we are considering only those shape invariant superpotentials that have at least one singularity at either the origin or $\infty$. If the superpotential has no poles at these extremes, then it must have poles at ``finite" fixed-points. These are potentials of trigonometric type,  and as noted earlier, will not be considered in this paper.

It is not hard to show that the case $b_{-1}=0, ~c_{-1}\neq0$ is identical to the case $b_{-1}\neq0,~c_{-1}=0$. Hence, we need consider only two cases: $b_{-1}\neq0, ~c_{-1}=0$ and $b_{-1}\neq0, ~c_{-1}\neq 0$. However, it turns out that for $\left( b_{-1}\neq0, ~c_{-1}\neq0\right)$, the quantization condition $\oint p\,dx = -2\pi i n$, does not yield an equation involving energy, and hence energy cannot be determined from singularities at boundary points alone. The QMF in such a case must have poles at points other than the origin and $\infty$ and, as stated earlier, such cases will not be considered here.  It is important to note that there are no exponential type known shape invariant potentials in quantum mechanics for which the superpotential has a singularities for both $x=\pm \infty$. Thus, in the following,  we will only consider the case where $b_{-1}\neq 0$ and $c_{-1} =  0$.

Near $r = 0$, the superpotential is given by
\begin{eqnarray} 
W(r, \alpha_0) = \frac{b_{-1}(\alpha_0)}{r} + b_0(\alpha_0) + b_1(\alpha_0)r +\cdots
\end{eqnarray} 
From $W$, we get
	$\frac{dW}{dx} = \frac{dW}{dr} \frac{dr}{dx} = r \frac{dW}{dr} =  - \frac{b_{-1}}{r}  +  b_1r$, and 	
	$W^2 = \frac{b_{-1}^2}{r^2} + \frac{2b_{-1}b_0}{ r} + b_0^2 + 2b_{-1}b_1 + \cdots$.
Hence,
due to the above change of variable, the shape invariance condition now takes the form
\begin{eqnarray} 
W^2(\alpha_0) + r \frac{dW(\alpha_0)}{dr} = W^2( \alpha_1 ) - r \frac{dW( \alpha_1 )}{dr} + R(\alpha_0). \label{ExpoSIC}
\end{eqnarray} 
Substituting the expansion of $W$ into Eq. (\ref{ExpoSIC}), and matching various powers of $r$, we get the following set of difference equations:
\begin{eqnarray} 
\begin{array}{ll}
{\rm from ~the}~r^{-2}~{\rm terms}  ~~~~~~~~ & b_{-1}^2(\alpha_0) = b_{-1}^2( \alpha_1 );\\
{\rm from ~the}~r^{-1}~{\rm terms}  & 2b_{-1}(\alpha_0)b_0(\alpha_0) - b_{-1} (\alpha_0)= 2b_{-1}(\alpha_1)b_0( \alpha_1 ) + b_{-1}(\alpha_1);\\
{\rm from ~the}~r^{0}~{\rm terms}  & b_0^2(\alpha_0) + 2b_{-1}(\alpha_0)b_1(\alpha_0) = b_0^2( \alpha_1 ) + 2b_{-1}(\alpha_1)b_1( \alpha_1 ) + R(\alpha_0).
\end{array}
\label{Expo-W3}
\end{eqnarray} 

From Eq. (\ref{Expo-W3}),  the first constraint implies that $b_{-1}(\alpha_0)$ does not depend upon its argument; i.e., it is a constant. We denote this constant by $b_{-1}$. The second constraint we get from matching the coefficients of $r^{-1}$ gives
$b_0(\alpha_1) = b_0( \alpha_0) -1$.
Substituting $\alpha_1  = \alpha_0 + 1$, we get the difference equation $$ b_0(\alpha_0+1) = b_0( \alpha_0) -1,$$ whose solution is $b_0(\alpha_0) =  - \alpha_0 + C,$ where $C$ is a constant. Now, from the last constraint of the set:~ $b_0^2(\alpha_0) + 2b_{-1}b_1(\alpha_0) = b_0^2( \alpha_1 ) + 2b_{-1}b_1( \alpha_1 ) + R(\alpha_0)$, we have
$$(- \alpha_0+C)^2 + 2b_{-1}b_1(\alpha_0) = ( -\alpha_1 + C)^2 + 2b_{-1}b_1( \alpha_1 ) + g( \alpha_1) - g(\alpha_0).$$
This equation can be written as
$$b_1(\alpha_0)+ \frac{g(\alpha_0)+ (- \alpha_0+C)^2}{2b_{-1}} = b_1(\alpha_1)+ \frac{g(\alpha_1)+ (- \alpha_1+C)^2}{2b_{-1}} ,$$
which shows that the expression $b_1(\alpha_0)+ \frac{g(\alpha_0)+ (- \alpha_0+C)^2}{2b_{-1}}$ is independent of its argument $\alpha_0$, hence we set it equal to a constant $D$. Thus, we get
\begin{eqnarray}  b_1(\alpha_0) = D - \left(  \frac{g(\alpha_0)+ (- \alpha_0+C)^2}{2b_{-1}}  \right). \end{eqnarray} 
We carry out a very similar analysis in the region $r\rightarrow \infty$, where the superpotential has the form given in Eq.  (\ref{exp_Wu}); i.e., $\widetilde{W}(u, \alpha_0) =  c_0 (\alpha_0)+ c_{1}(\alpha_0)\,u + c_{1}(\alpha_0)\,u^2 \cdots $, and we get
$$\left(  c_{0}(\alpha_0) \right)^2 = \left(  c_{0}(\alpha_1) \right)^2 + R(\alpha_0)$$
Writing this equation as $\left(  c_{0}(\alpha_0) \right)^2 + g(\alpha_0)  = \left(  c_{0}(\alpha_1) \right)^2 + g(\alpha_1)$, we see that the expression  $\left(  c_{0}(\alpha_0) \right)^2 + g(\alpha_0) $ is independent of $\alpha_0$. We set it equal to a constant $\Theta$. Thus, we get
\begin{eqnarray} 
c_{0}(\alpha_0)  =  \pm \sqrt{\Theta - g(\alpha_0)}  . \nonumber
\end{eqnarray} 
Thus, to summarize, the structure of  $W$ at two end points are given by
$$
W(r,\alpha_0) = \frac{b_{-1}}{r} + (- \alpha_0 + C) + \left[  D - \left(  \frac{g(\alpha_0)+ (- \alpha_0+C)^2}{2b_{-1}}  \right)\right]\, r +\cdots~,
$$
and
$$\widetilde{W}(u, \alpha_0) =  \left(  \pm \sqrt{\Theta - g(\alpha_0)} \right) + c_{1}(\alpha_0)\,u + c_{1}(\alpha_0)\,u^2 \cdots .$$

We will now substitute these two forms of the superpotential into the QHJ equation to determine the analytic structure of $p$ and from it the eigenvalues of the shape invariant system.

Near $r=0$, the QHJ equation is given by
\begin{eqnarray} 
p^2-r \frac{dp}{dr} = W^2(r,\alpha_0) - r \frac{dW(r,\alpha_0)}{dr} -E, \label{ExpoQHJ-origin}
\end{eqnarray} 
where $W(r,\alpha_0)$ is given by
\mbox{$W(r,\alpha_0) = \frac{b_{-1}}{r} + (- \alpha_0 + C) + \left[  D - \left(  \frac{g(\alpha_0)+ (- \alpha_0+C)^2}{2b_{-1}}  \right)\right]\, r +\cdots
.$}
Expanding the quantum momentum function as $p(r)=p_{-1}r^{-1} + p_0 + p_1r + \cdots$, and substituting it into Eq. (\ref{ExpoQHJ-origin}), we get
\begin{eqnarray} 
p_{-1}^2 = b_{-1}^2;~~~~2p_{-1}p_0 - p_{-1} = 2b_{-1}b_0(\alpha_0) - b_{-1}; ~~~{\rm etc.}
\end{eqnarray} 
From the first equality in above equation, we get $p_{-1} = \pm b_{-1}$. However, in the limit $E\to 0$, we must have $p \to W$.  This implies  $p_{-1} = b_{-1}$. Substituting this value of $p_{-1}$ in the second equality, we get  $p_0 = b_0(\alpha_0) = -\alpha_0 + C.$ As we will soon see, we need go no further.  $p_0$  is all we need to determine the contour integral around $r=0$.

Near $u= 0$, i.e., $r\to \infty$, we expand the QMF as $\widetilde{p}(r)=\widetilde{p}_{-1}u^{-1} + \widetilde{p}_0 + \widetilde{p}_1u + \cdots$. We substitute this expansion for $\widetilde{p}(r)$ and the superpotential 
\mbox{$\widetilde{W}(u, \alpha_0) =  \left(  \pm \sqrt{\Theta - g(\alpha_0)} \right) + c_{1}(\alpha_0)\,u + c_{1}(\alpha_0)\,u^2 \cdots $}
into the QHJ equation
\begin{eqnarray} 
 \widetilde{p}^2+u \frac{d \widetilde{p}}{du} =  \widetilde{W}^2(\alpha_0) + u \frac{d \widetilde{W}(\alpha_0)}{du} -E.
\end{eqnarray} 
Now comparing various powers of $u$, we get $\widetilde{p}_{-1} =0$, $\left(\widetilde{p}_{0}\right)^2 =
  \left(  \pm \sqrt{\Theta - g(\alpha_0)} \right)^2 - E$, etc.  Thus, we get $\widetilde{p}_{0} =
\pm \sqrt{\Theta - g(\alpha_0)-E}$.

The quantization condition in this case is
\begin{eqnarray} 
\oint_{x=-\infty} p ~dx  + \oint_{x=\infty} p ~dx  = ~2\pi \,i\,n~; ~~~~{\rm i.e.},  \nonumber\\\nonumber\\
\oint_{r=0} \left(\frac{p_0}{r}  \right) dr  - \oint_{u=0} \left(  \frac{\widetilde{p}_0}{u}  \right) du = ~2\pi \,i\,n~;  \nonumber\\\nonumber\\
(C-\alpha_0) \pm \sqrt{\Theta - g(\alpha_0)-E_n} =~n~\nonumber  ,
\end{eqnarray} 
which yields,
\begin{eqnarray} 
E_n = \Theta- \left(C-\alpha_0 -n  \right)^2 - g(\alpha_0) \nonumber
\end{eqnarray} 
Since we assume that SUSY is unbroken, we must have $E_0=0$. This implies $ g(\alpha_0)  = \Theta- \left(C-\alpha_0   \right)^2$, and hence, $ g(\alpha_n)  = \Theta- \left(C-(\alpha_0+n)   \right)^2 =  \Theta- \left(C-\alpha_0 -n  \right)^2$.
Thus, we have
\begin{eqnarray} 
E_n = \left(C-\alpha_0   \right)^2 - \left( C-\alpha_0-n   \right)^2 = g(\alpha_n) - g(\alpha_0) ~.
\end{eqnarray} 
If we identify $C-\alpha_0$ with the parameter $A$ of Morse superpotential $W = A- Be^{-x}$, we get the energy $E_n = A^2-(A-n)^2$, which is exactly what is expected for the Morse potential.

\vspace*{.2in}
\noindent
{\bf Conclusion:}
One of the current authors (AG) and his collaborators had previously connected Quantum Hamilton-Jacobi Theory with supersymmetric quantum mechanics, and had shown that the quantum momenta for a shape invariant system can be derived recursively, analogous to SUSYQM. In this paper, we have further explored the impact of shape invariance on Quantum Hamilton-Jacobi theory. In particular, we showed that the shape invariance condition provides sufficient information about the singularity structure of QMF's to determine the eigenvalues of the system.

There are several directions in which this work could be extended. We only considered translational shape invariant potentials; that is, those for which partner potentials are connected by a shift in parameter ($a_{i+1} = a_i+1$). This leaves out the very important case where parameters in partner potentials are multiplicatively connected, i.e., $a_{i+1} = q a_i$, where $q \in (0,1)$. In addition, we only considered the potentials for which all fixed poles were at the origin or at $\infty$ or both. Potentials with finite domain, and those for which the superpotential has poles at points other than $0$ or/and  $\infty$ can also be explored.  Recently, the work in SUSYQM has been extended to fractional supersymmetric quantum mechanics \cite{Fsusy1,Fsusy2}. One of the benefits of this formalism is that several solvable potentials are generated from one generalized potential for varying values of a parameter. The solvability in this case is described by a generalized shape invariance condition that is somewhat more involved than the additive shape invariance considered here. It would be worth exploring if the present formalism can be extended to fractional SUSYQM.

We would like to thank Prof. John Dykla,  Prof. Constantine Rasinariu and Walter Moore for discussions.  We would also like to thank Prof.  Jeffry Mallow for a careful reading of the manuscripts and providing many helpful suggestions. We would also like to thank one of the referees for several helpful suggestions.

\end{document}